\documentclass[journal]{IEEEtran}
\usepackage{mathtools}

\usepackage{fancyhdr}
\usepackage{amsmath,amssymb}
\usepackage{amsfonts}
\usepackage{graphicx}
\usepackage{dsfont}
\usepackage{comment}
\usepackage{algorithmic}
\usepackage{comment}
\usepackage{caption}
\usepackage{alphabeta}
\usepackage{balance}
\usepackage{algorithm}
\usepackage{float}
\usepackage[hidelinks]{hyperref}    %for using hyperlinks
\usepackage{color}

\usepackage[caption=false,font=footnotesize]{subfig}
\usepackage{pgfplots}
\pgfplotsset{compat=1.18}
\usepackage{enumerate}
\usepackage{graphics}
\usepackage{cite}
\usepackage{mathrsfs}
\usepackage{stfloats}

\usepackage{tikz}
\usepackage{mathdots}
\usepackage{yhmath}
\usepackage{cancel}
\usepackage{color}
\usepackage{siunitx}
\usepackage{array}
\usepackage{multirow}
\usepackage{textcomp}
\usepackage{gensymb}
\usepackage{tabularx}
\usepackage{booktabs}
\usetikzlibrary{fadings}
\usetikzlibrary{patterns}
\usetikzlibrary{shadows.blur}
\usepackage{graphicx}

\begin{document}

\title{Elliptic Range-Doppler Mapping for OFDM-ISAC under IQ Imbalance}
\author{Thrassos K. Oikonomou, Dimitrios Bozanis, Sotiris A. Tegos, \IEEEmembership{Senior Member, IEEE,}
Panagiotis D. Diamantoulakis,  
\IEEEmembership{Senior Member, IEEE,} and George K. Karagiannidis,~\IEEEmembership{Fellow, IEEE}
\thanks{T. K. Oikonomou, D. Bozanis, S. A. Tegos, P. D. Diamantoulakis, and G. K. Karagiannidis are with the Department of Electrical and Computer Engineering, Aristotle University of Thessaloniki, Thessaloniki, Greece (e-mails: \{toikonom, dimimpoz, tegosoti, padiaman, geokarag\}@ece.auth.gr)}
\vspace{-3mm}
}
\maketitle

\begin{abstract} 
Receiver in-phase/quadrature imbalance (IQI) couples each OFDM subcarrier with its mirror counterpart, creating ghost targets and degrading range-Doppler recovery in orthogonal frequency division multiplexing (OFDM) integrated sensing and communication (ISAC). Instead of first compensating for IQI and then applying conventional processing, this letter exploits the structure of the IQI-impaired observation directly. We show that each physical target induces coupled direct and mirror components linked through the target coefficient and its conjugate, which motivates an elliptic atom group representation for each candidate delay-Doppler cell. Based on this model, we propose an elliptic group orthogonal matching pursuit detector that performs sparse recovery directly on the received OFDM grid. The required correlations are computed efficiently through two weighted two-dimensional fast Fourier transforms (FFTs) followed by local group projections. Numerical results show that the proposed method improves exact support recovery and weak-target detection compared to corresponding benchmarks,
 % OFDM-ZF+OMP, Pairwise IQI-ZF+OMP, and LMMSE+OMP, 
especially under moderate and strong receiver IQI.
\end{abstract}


\section{Introduction}
Integrated sensing and communication (ISAC) has emerged as a key paradigm for sixth-generation (6G) wireless networks, where the same spectrum, hardware, and waveform resources are reused for both data transmission and environmental sensing \cite{isac_1}. From a signal-processing perspective, orthogonal frequency division multiplexing (OFDM) is particularly attractive for communication-centric ISAC because it is already adopted in modern wireless standards and its frequency-domain structure naturally supports range-Doppler processing \cite{isac_2}, \cite{isac_3}. In the ideal transceiver case, the received OFDM resource grid can be converted into a channel matrix and processed by Fourier operations across subcarriers and OFDM symbols to obtain a range-Doppler map.

Several works have investigated OFDM-based sensing and its practical impairments. Classical OFDM radar processing and its range-Doppler interpretation were developed in early radar-communication studies \cite{isac_3}, \cite{sota_1}. More recently, the impact of in-phase/quadrature imbalance (IQI) on OFDM-ISAC has received attention, since in-phase/quadrature (IQ) mismatch introduces mirror-subcarrier interference, ghost targets, and an increased range-Doppler floor. IQI-robust OFDM radar waveforms were proposed in \cite{sota_2}, while OFDM-ISAC waveforms robust to frequency-selective-IQI and ghost-relocation strategies were developed in \cite{sota_3}. Compensation-based approaches have also been proposed, including transmitter and receiver IQI estimation and compensation for 5G NR-compliant ISAC before standard OFDM radar processing \cite{sota_4}. In parallel, sparse and super-resolution delay-Doppler processing for OFDM radar has been studied using compressed sensing and atomic-norm methods \cite{sota_5}, and recent works have examined the impact of random communication symbols and sidelobes in communication-centric OFDM-ISAC \cite{sota_6}, \cite{sota_7}.

Despite these advances, existing IQI-aware OFDM sensing methods mainly address IQI through explicit compensation before conventional range-Doppler processing. Such approaches are effective when the goal is to restore an ordinary OFDM sensing model, but do not exploit the structure of the IQI-impaired observation itself. In particular, receiver IQI does not create an arbitrary distortion in the range-Doppler domain. Instead, a single physical target gives rise to two coupled components, a direct branch at the target delay-Doppler cell and a mirror-conjugate branch with the same delay and opposite Doppler. These two components are linked through the same complex target coefficient and its conjugate. Therefore, the IQI-impaired response of one physical target is no longer well described by a single ordinary delay-Doppler atom, but by a structured two-branch atom. This observation naturally motivates an elliptic target representation, in which the direct and mirror components are treated jointly as one physical range-Doppler signature. Although sparse OFDM sensing methods have been studied using ordinary delay-Doppler dictionaries, they do not model this IQI-induced mirror-conjugate coupling. To the best of our knowledge, direct sparse range-Doppler recovery using receiver-IQI-aware elliptic target atoms has not previously been investigated.

In this paper, we propose an elliptic range-Doppler processing framework for OFDM-ISAC under receiver IQI. First, we derive the IQI-impaired OFDM sensing model and show that each physical target induces a direct-plus-mirror elliptic atom group rather than an isolated ordinary delay-Doppler atom. Second, we formulate an elliptic group orthogonal matching pursuit (GOMP) detector that performs sparse target recovery directly on the IQI-impaired observations, without first reconstructing a compensated channel. Third, we show that the required elliptic correlations can be computed efficiently using two weighted two-dimensional fast Fourier transforms (FFTs) followed by a local two-dimensional projection. Finally, numerical results demonstrate that the proposed method improves exact support recovery and ghost suppression compared to conventional orthogonal matching pursuit (OMP), ZF-compensated OMP, and LMMSE-compensated OMP baselines, especially in moderate and strong-IQI, multi-target, and weak-target regimes.

\section{System Model}
We consider an OFDM-ISAC frame consisting of \(M\) OFDM symbols and \(N\) subcarriers. The transmitted data symbol on the \((m,k)\)-th resource element is denoted by \(X_{m,k}\), where \(m=0,\ldots,M-1\) and \(k=0,\ldots,N-1\). The transmitted grid is assumed to be known at the sensing receiver and normalized such that the average symbol power is given by $\mathbb{E}\left[|X_{m,k}|^2\right]=1$.
The sensing scene is modeled as a sparse collection of \(K\) point targets in the delay-Doppler domain. Throughout this letter, we adopt an on-grid delay-Doppler model. Specifically, the frequency-domain channel over the OFDM grid is written as
\begin{equation}
H_{m,k}
=
\sum_{p=1}^{K}
\alpha_p
e^{-j2\pi k\tau_p/N}
e^{j2\pi m\nu_p/M},
\label{eq:channel_model}
\end{equation}
where \(\alpha_p\in\mathbb{C}\) denotes the complex reflection coefficient of the \(p\)-th target, while \(\tau_p=\{0,\dots,N-1\}\) and \(\nu_p=\{0,\dots,N-1\}\) denote its delay and Doppler indices, respectively. The set of active delay-Doppler target locations is denoted by $\mathcal{S}=\{(\tau_p,\nu_p)\}_{p=1}^{K}$.
% \begin{equation}
% \mathcal{S}
% =
% \{(\tau_p,\nu_p)\}_{p=1}^{K}.
% \label{eq:support_set}
% \end{equation}

In the absence of receiver hardware impairments, the received OFDM sensing grid is given by
\begin{equation}
R_{m,k}
=
X_{m,k}H_{m,k}
+
W_{m,k},
\label{eq:ideal_rx}
\end{equation}
where \(W_{m,k}\sim\mathcal{CN}(0,\sigma_w^2)\) is additive noise. Moreover, the standard delay-Doppler atom associated with a candidate cell \(\theta=(\tau,\nu)\) is defined as
\begin{equation}
\phi_{\theta}[m,k]
=
e^{-j2\pi k\tau/N}
e^{j2\pi m\nu/M}.
\label{eq:standard_atom}
\end{equation}
Accordingly, the ideal channel can be expressed as
\begin{equation}
H
=
\sum_{p=1}^{K}
\alpha_p
\phi_{\tau_p,\nu_p}.
\label{eq:ideal_sparse_channel}
\end{equation}

In this context, conventional sparse range-Doppler recovery relies on the assumption that each physical target is represented by a single delay-Doppler atom. However, this representation is altered in the presence of receiver IQI. The receiver IQI model is written in complex baseband as in \cite{rx_iqi}
\begin{equation}
x_{\mathrm{IQ}}(t)
=
\mu x(t)
+
\lambda x^*(t),
\label{eq:iqi_baseband}
\end{equation}
where \(\mu\) and \(\lambda\) are the direct and image IQI coefficients, respectively. It should be noted that frequency-flat IQI coefficients are considered. Let the mirror subcarrier index be given by $q(k)=(-k)\bmod N$. For even \(N\), the subcarrier \(k=0\) and the Nyquist subcarrier \(k=N/2\) satisfy \(q(k)=k\), and are therefore self-mirror bins. The model in in \eqref{eq:iqi_grid} remains valid for these bins, while pairwise IQI processing treats them as self-pairs.
Then, the receiver-IQI-impaired OFDM sensing grid is written as in \cite{rx_iqi}
\begin{equation}
Y_{m,k}
=
\mu X_{m,k}H_{m,k}
+
\lambda X_{m,q(k)}^*H_{m,q(k)}^*
+
W_{m,k}.
\label{eq:iqi_grid}
\end{equation}
The objective is to estimate the active delay-Doppler target locations \(\mathcal{S}\) from \(Y\), given the transmitted grid \(X\) and the IQI coefficients. In the rest of this work, the IQI coefficients are assumed to be estimated and available at the receiver. Therefore, we focus on the target detection problem conditioned on the available IQI information.

\section{OFDM-ISAC under Receiver IQI}
The IQI-impaired model in \eqref{eq:iqi_grid} shows that receiver IQI changes the effective delay-Doppler signature observed by the sensing receiver. Specifically, receiver IQI couples each subcarrier with its mirror counterpart, so that a physical target no longer appears only through the standard OFDM sensing atom. To make this structure explicit, consider a single target located at \((\tau,\nu)\). Its ideal channel response is given by
\begin{equation}
H_{m,k}
=
\alpha
e^{-j2\pi k\tau/N}
e^{j2\pi m\nu/M}.
\label{eq:single_target_channel}
\end{equation}
The corresponding mirror-conjugate channel term appearing in the receiver-IQI-impaired observation is written as
\begin{equation}
H_{m,q(k)}^*
=
\alpha^*
e^{-j2\pi k\tau/N}
e^{-j2\pi m\nu/M}.
\label{eq:single_target_mirror}
\end{equation}
Therefore, the IQI-induced image preserves the delay phase of the physical target but reverses its Doppler phase. In more detail, the contribution of this target to the impaired OFDM grid is given by
\begin{equation}
\begin{aligned}
Y^{(\tau,\nu)}_{m,k}
=
&\ \mu \alpha X_{m,k}
e^{-j2\pi k\tau/N}
e^{j2\pi m\nu/M}
\\
&+
\lambda \alpha^* X_{m,q(k)}^*
e^{-j2\pi k\tau/N}
e^{-j2\pi m\nu/M}.
\end{aligned}
\label{eq:single_target_iqi_contribution}
\end{equation}
Thus, receiver IQI produces a direct branch associated with \((\tau,\nu)\) and a mirror branch associated with \((\tau,\bar{\nu})\), where $\bar{\nu}=-\nu\bmod M$. These two branches are not independent scatterers, since they are coupled through the same reflection coefficient \(\alpha\) and its conjugate \(\alpha^*\).

The most common OFDM sensing front-end first removes the known transmitted symbols through element-wise division. The resulting channel estimate, denoted here as OFDM-ZF, is given by
\begin{equation}
\widetilde H_{m,k}^{\mathrm{ZF}}
=
\frac{Y_{m,k}}{X_{m,k}}.
\label{eq:ofdm_zf}
\end{equation}
This operation is zero-forcing only with respect to the transmitted OFDM symbols and does not account for receiver IQI. For the single target above, the OFDM-ZF output is written as
\begin{equation}
\begin{aligned}
\widetilde H_{\mathrm{ZF}}^{(\tau,\nu)}[m,k]
=
&\ \mu \alpha
e^{-j2\pi k\tau/N}
e^{j2\pi m\nu/M}
\\
&+
\lambda \alpha^*
d_{m,k}
e^{-j2\pi k\tau/N}
e^{-j2\pi m\nu/M},
\end{aligned}
\label{eq:ofdm_zf_single_target}
\end{equation}
where the data-dependent mirror weighting term is defined as $d_{m,k}=X_{m,q(k)}^*/X_{m,k}$.
% \begin{equation}
% d_{m,k}
% =
% \frac{X_{m,q(k)}^*}{X_{m,k}}.
% \label{eq:data_dependent_weight}
% \end{equation}
The first term in \eqref{eq:ofdm_zf_single_target} remains a standard delay-Doppler atom. In contrast, the second term has the opposite Doppler phase but is additionally modulated by \(d_{m,k}\). Thus, after OFDM-ZF channel formation, the IQI image is generally not a clean Fourier atom. Instead, it becomes a data-weighted mirror response, which can raise the range-Doppler floor, create random-symbol-induced artifacts, and increase the probability of selecting ghost locations \cite{sota_4}.

A natural next step is to account for receiver IQI explicitly before range-Doppler recovery. This can be done by grouping each subcarrier with its mirror counterpart. Specifically, the received vector is defined as
\begin{equation}
\mathbf y_{m,k}
=
\begin{bmatrix}
Y_{m,k}\\
Y_{m,q(k)}^*
\end{bmatrix},
\label{eq:pairwise_y}
\end{equation}
and the corresponding channel vector is defined as
\begin{equation}
\mathbf h_{m,k}
=
\begin{bmatrix}
H_{m,k}\\
H_{m,q(k)}^*
\end{bmatrix}.
\label{eq:pairwise_h}
\end{equation}
The pairwise receiver IQI model is then written as
\begin{equation}
\mathbf y_{m,k}
=
A_{m,k}\mathbf h_{m,k}
+
\mathbf w_{m,k},
\label{eq:pairwise_model}
\end{equation}
where the mixing matrix is given by
\begin{equation}
A_{m,k}
=
\begin{bmatrix}
\mu X_{m,k} & \lambda X_{m,q(k)}^*\\
\lambda^*X_{m,k} & \mu^*X_{m,q(k)}^*
\end{bmatrix}.
\label{eq:pairwise_matrix}
\end{equation}
Direct inversion of this system yields the IQI-suppressed estimate, denoted here as pairwise IQI-ZF, which is written as in \cite{rx_iqi}
\begin{equation}
\widehat{\mathbf h}^{\mathrm{IQI-ZF}}_{m,k}
=
A_{m,k}^{-1}
\mathbf y_{m,k}.
\label{eq:pairwise_iqi_zf}
\end{equation}
Unlike OFDM-ZF, pairwise IQI-ZF explicitly accounts for the receiver IQI coupling. However, its robustness depends on the conditioning of \(A_{m,k}\). Specifically, the determinant of the mixing matrix is given by $\det(A_{m,k})=X_{m,k}X_{m,q(k)}^*\left(|\mu|^2 - |\lambda|^2\right)$.
% \begin{equation}
% \det(A_{m,k})
% =
% X_{m,k}X_{m,q(k)}^*
% \left(|\mu|^2-|\lambda|^2\right).
% \label{eq:pairwise_det}
% \end{equation}
This expression shows that direct IQI inversion can become noise-sensitive when the IQI level is strong, namely when \(|\lambda|\) approaches \(|\mu|\), or when high-order QAM symbols produce small values of \(|X_{m,k}|\) or \(|X_{m,q(k)}|\). Therefore, while pairwise IQI-ZF suppresses the receiver IQI image, it may also amplify noise before range-Doppler recovery.

To mitigate this noise enhancement, one may regularize the pairwise inversion. The corresponding linear minimum mean-square error estimate, denoted here as LMMSE, is written as
\begin{equation}
\widehat{\mathbf h}^{\mathrm{LMMSE}}_{m,k}
=
\left(
A_{m,k}^H A_{m,k}
+
\gamma I_2
\right)^{-1}
A_{m,k}^H
\mathbf y_{m,k},
\label{eq:lmmse_estimator}
\end{equation}
where \(\gamma\) is a regularization parameter related to the noise and channel powers, and $I_2$ denotes the $2\times2$ identity matrix. This regularization improves stability compared to direct inversion and limits noise amplification. Nevertheless, both pairwise IQI-ZF and LMMSE follow the same IQI-suppression-first philosophy, where a conventional channel estimate is reconstructed before range-Doppler recovery.

This compensation-first strategy can be effective when the IQI parameters are accurately known and the inversion is well conditioned. However, it does not directly exploit the fact that the direct and mirror branches are generated by the same physical target and are linked through \(\alpha\) and \(\alpha^*\). Moreover, the suppression stage may amplify noise, color the residual disturbance, or distort weak target responses before detection. This motivates a different strategy, where instead of first converting the received grid into a conventional single-branch channel estimate, the sensing receiver should operate directly on the IQI-impaired observation and jointly match the direct and mirror branches of each physical target. The next section develops this idea through an elliptic target representation and a fast group sparse recovery method.

\section{Fast Elliptic GOMP Detector}

Conventional OFDM range-Doppler recovery represents a target at \(\theta=(\tau,\nu)\) through a single Fourier atom, namely \(H_{\theta}=\alpha\phi_{\theta}\). This single-atom model is no longer complete under receiver IQI, because the received response of one physical target contains both a direct branch and a mirror-conjugate branch. For a candidate cell \(\theta=(\tau,\nu)\), these two branches are given by
\begin{equation}
\begin{aligned}
a_{\theta}[m,k]
&=
\mu X_{m,k}
e^{-j2\pi k\tau/N}
e^{j2\pi m\nu/M},\\
b_{\theta}[m,k]
&=
\lambda X_{m,q(k)}^*
e^{-j2\pi k\tau/N}
e^{-j2\pi m\nu/M}.
\end{aligned}
\label{eq:direct_mirror_branches}
\end{equation}
Thus, the IQI-impaired response of one physical target can be written as
\begin{equation}
Y_{\theta}
=
\alpha a_{\theta}
+
\alpha^*b_{\theta}.
\label{eq:iqi_target_response}
\end{equation}
This expression shows that the direct and mirror branches are not independent scatterers, since they are controlled by the same reflection coefficient \(\alpha\) and its conjugate \(\alpha^*\).
Writing \(\alpha=\alpha_R+j\alpha_I\), the response in \eqref{eq:iqi_target_response} can be expressed as $\alpha a_{\theta}
+
\alpha^*b_{\theta}
=
\alpha_R g_{1,\theta}
+
\alpha_I g_{2,\theta},$
where the two real axes of the target response are defined as $g_{1,\theta}=a_{\theta}+b_{\theta}$, $g_{2,\theta}=j(a_{\theta}-b_{\theta}).$

Consequently, the IQI-aware target atom is the real two-dimensional group \(G_{\theta}=[g_{1,\theta},g_{2,\theta}]\) and the contribution of one target is compactly written as
\begin{equation}
Y_{\theta}
=
G_{\theta}\beta_{\theta},
\quad
\beta_{\theta}
=
[\alpha_R,\alpha_I]^T,
\label{eq:elliptic_group_model}
\end{equation}
where $\alpha_R,\alpha_I\in \mathbb{R}$.
% This construction explains the term elliptic atom. For a unit-magnitude coefficient \(\alpha=e^{j\varphi}\), the target response is given by
% \begin{equation}
% Y_{\theta}(\varphi)
% =
% g_{1,\theta}\cos\varphi
% +
% g_{2,\theta}\sin\varphi.
% \label{eq:elliptic_locus_compact}
% \end{equation}
% Thus, as the unknown target phase varies, the possible received responses are traced by the two real axes \(g_{1,\theta}\) and \(g_{2,\theta}\). Without IQI, this reduces to the circular phase rotation of a conventional Fourier atom. With IQI, the two axes generally have unequal norms and nonzero real inner product, so the phase trajectory is deformed into an ellipse in the real observation space. The proposed elliptic atom therefore generalizes the conventional Fourier atom by representing the complete direct-plus-mirror signature of one physical target.

This construction explains both the term elliptic atom and the associated elliptic subspace. Let the complex target coefficient be written as \(\alpha=\rho e^{j\varphi}\), where \(\rho\geq 0\) denotes the target amplitude and \(\varphi\) denotes its phase. Then, the IQI-impaired response of a target at the candidate cell \(\theta\) is given by
\begin{equation}
Y_{\theta}(\rho,\varphi)
=
\rho
\left(
g_{1,\theta}\cos\varphi
+
g_{2,\theta}\sin\varphi
\right).
\label{eq:elliptic_locus_compact}
\end{equation}
For a fixed amplitude \(\rho\), the unknown target phase traces a closed curve generated by the two real axes \(g_{1,\theta}\) and \(g_{2,\theta}\). Without IQI, this curve reduces to the circular phase rotation of a conventional Fourier atom. However, with IQI, the two axes generally have unequal norms and nonzero real inner product, so the fixed-amplitude phase trajectory is deformed into an ellipse in the real observation space.
Since the target amplitude is also unknown, the admissible responses are not restricted to a single ellipse. Instead, they consist of all scaled versions of this ellipse, which form the real two-dimensional subspace $\mathcal{U}_{\theta}=\operatorname{span}_{\mathbb{R}}
\{g_{1,\theta},g_{2,\theta}\}$.
% \begin{equation}
% \mathcal{U}_{\theta}
% =
% \operatorname{span}_{\mathbb{R}}
% \{g_{1,\theta},g_{2,\theta}\}.
% \label{eq:elliptic_subspace}
% \end{equation}
Equivalently, every response in \(\mathcal{U}_{\theta}\) can be written as \(G_{\theta}\beta_{\theta}\), where \(G_{\theta}=[g_{1,\theta},g_{2,\theta}]\) and \(\beta_{\theta}=[\alpha_R,\alpha_I]^T\). The proposed elliptic atom therefore generalizes the conventional Fourier atom by replacing the circular response family of a single direct branch with an IQI-aware elliptic subspace that represents the complete direct-plus-mirror signature of one physical target.

Let \(y=\operatorname{vec}(Y)\). Using the elliptic atom groups, the IQI-impaired sparse sensing model is written as
\begin{equation}
y
=
\sum_{\theta\in\mathcal{S}}
G_{\theta}\beta_{\theta}
+
w,
\label{eq:elliptic_sparse_model_compact}
\end{equation}
and the corresponding recovery problem is given by
\begin{equation}
\min_{\mathcal{S},\{\beta_{\theta}\}}
\left\|
y-
\sum_{\theta\in\mathcal{S}}
G_{\theta}\beta_{\theta}
\right\|_2^2,
\qquad
|\mathcal{S}|=K.
\label{eq:elliptic_sparse_problem_compact}
\end{equation}
Unlike OFDM-ZF, pairwise IQI-ZF, and LMMSE processing, this formulation does not first reconstruct a conventional single-branch channel estimate. Instead, it directly fits the IQI-impaired observation using target groups that contain both the direct and mirror branches.

To solve \eqref{eq:elliptic_sparse_problem_compact} efficiently, we propose an elliptic GOMP. Following the greedy structure of OMP, the method replaces scalar Fourier-atom matching with a group projection onto the real two-dimensional elliptic subspace of each candidate delay-Doppler cell. At each iteration, it selects the cell whose elliptic group captures the largest residual energy. In this way, the algorithm searches for complete physical target responses rather than isolated direct or mirror components. Let \(r\) denote the current vector residual and define
\begin{equation}
\bar r=
[\Re\{r\}^T,\Im\{r\}^T]^T,
\quad
\bar G_{\theta}
=
\begin{bmatrix}
\Re\{g_{1,\theta}\} & \Re\{g_{2,\theta}\}\\
\Im\{g_{1,\theta}\} & \Im\{g_{2,\theta}\}
\end{bmatrix}.
\label{eq:real_projection_definitions}
\end{equation}
The elliptic projection score is given by
\begin{equation}
T_E(\theta)
=
\bar r^T
\bar G_{\theta}
\left(
\bar G_{\theta}^T
\bar G_{\theta}
\right)^{-1}
\bar G_{\theta}^T
\bar r.
\label{eq:elliptic_projection_score_compact}
\end{equation}
The selected cell is then obtained as \(\widehat{\theta}=\arg\max_{\theta}T_E(\theta)\). After each selection, the coefficient vector is updated by least squares over the selected groups, while the residual is updated by subtracting the reconstructed direct-plus-mirror responses. In particular, for the selected set \(\widehat{\mathcal{S}}_i\), the least-squares update is written as
\begin{equation}
\widehat{\beta}_{\widehat{\mathcal{S}}_i}
=
\arg\min_{\beta}
\left\|
\bar y
-
\bar G_{\widehat{\mathcal{S}}_i}\beta
\right\|_2^2,\hspace{0.2cm}
\bar y=
[\Re\{y\}^T,\Im\{y\}^T]^T .
\label{eq:group_ls_compact}
\end{equation}
This update removes both branches of each detected target, which is essential when weak targets are masked by the IQI-induced response of stronger ones.

A direct evaluation of \eqref{eq:elliptic_projection_score_compact} over all candidate cells would require explicitly constructing the elliptic dictionary. To avoid this, the required inner products are computed through weighted two-dimensional Fourier operations. Let \(R\in\mathbb{C}^{M\times N}\) denote the current residual reshaped on the OFDM grid. The direct and mirror branch correlations are given by
\begin{equation}
\begin{aligned}
C_a[\tau,\nu]
&=
\sum_{m,k}
R_{m,k}
\mu^*X_{m,k}^*
e^{j2\pi k\tau/N}
e^{-j2\pi m\nu/M},\\
C_b[\tau,\nu]
&=
\sum_{m,k}
R_{m,k}
\lambda^*X_{m,q(k)}
e^{j2\pi k\tau/N}
e^{j2\pi m\nu/M}.
\end{aligned}
\label{eq:fast_branch_correlations}
\end{equation}
These quantities are evaluated using two weighted two-dimensional FFTs. Moreover, for \(\theta=(\tau,\nu)\), they satisfy \(g_{1,\theta}^Hr=C_a[\theta]+C_b[\theta]\) and \(g_{2,\theta}^Hr=-j(C_a[\theta]-C_b[\theta])\). Therefore, the real-valued correlation vector required for the projection is written as
\begin{equation}
c_{\theta}
=
\begin{bmatrix}
\Re\{C_a[\theta]+C_b[\theta]\}\\
\Re\{-j(C_a[\theta]-C_b[\theta])\}
\end{bmatrix}.
\label{eq:fast_c_theta_compact}
\end{equation}
With the local Gram matrix \(Q_{\theta}=\bar G_{\theta}^T\bar G_{\theta}\), the score in \eqref{eq:elliptic_projection_score_compact} is equivalently computed as \(T_E(\theta)=c_{\theta}^TQ_{\theta}^{-1}c_{\theta}\).
Since the common delay phase cancels in the inner products defining \(Q_{\theta}\), for \(\theta=(\tau,\nu)\), the Gram matrix depends only on the Doppler index, i.e., \(Q_{\theta}=Q_{\nu}\), and therefore at most \(M\) distinct \(2\times2\) Gram matrices are precomputed per OFDM frame.
Consequently, for \(K\) detected targets, the dominant complexity is due to the two weighted two-dimensional FFTs per greedy iteration and scales as \(\mathcal{O}\left(KMN\log(MN)\right)\), preserving the leading FFT order of conventional sparse OFDM range-Doppler processing while exploiting the receiver-IQI-induced elliptic geometry.
\vspace{-0.6cm}
\section{Numerical Results}
We consider an OFDM-ISAC frame with \(M=16\) OFDM symbols and \(N=64\) subcarriers, so that the delay-Doppler grid contains \(MN=1024\) candidate cells. The receiver IQI coefficients are \(\mu\),\(\lambda\) vary, and each point is averaged over \(10^3\) Monte Carlo trials. We use \(P_D\) to denote the probability of exact recovery of the active delay-Doppler target locations.

In Fig.~1, \(P_D\) is plotted versus SNR for \(K=10\) targets under moderate and strong receiver IQI, with \(|\lambda/\mu|=0.5\) in Fig.~1(a) and \(|\lambda/\mu|=0.8\) in Fig.~1(b). In both cases, the proposed elliptic GOMP reaches reliable exact recovery at lower SNR than the reference receivers. The gain is especially clear under strong IQI, where OFDM-ZF+OMP and pairwise IQI-ZF+OMP require much higher SNR, while LMMSE+OMP remains more robust but still lags behind the proposed method. This shows that jointly matching the direct and mirror target branches provides an effective sensing gain, since the IQI-induced image is exploited as part of the elliptic target signature rather than treated only as an impairment.

In Fig.~2, \(P_D\) is shown versus the number of targets for \(|\lambda/\mu|=0.6\) and SNR \(=0\) dB. When \(K\) is small, all methods recover the scene accurately because the delay-Doppler grid has \(1024\) candidate cells and the scene is highly sparse. As \(K\) increases, the mismatched and compensation-first receivers degrade rapidly due to residual IQI artifacts, noise enhancement, and wrong atom selections. LMMSE+OMP is more stable than OFDM-ZF+OMP and pairwise IQI-ZF+OMP, but it also loses reliability as the target density grows. In contrast, elliptic GOMP maintains high exact recovery across the full range of \(K\), showing that modeling and subtracting the complete direct-plus-mirror target response prevents IQI-induced ghosts from accumulating in denser scenes.

In Fig.~3, we evaluate a near-target scenario with \(K=2\), where a weak target is placed in the delay cell adjacent to a strong target and at the same Doppler index. The weak target is \(25\) dB below the strong one. Fig.~3(a) and Fig.~3(b) correspond to \(|\lambda/\mu|=0.5\) and \(|\lambda/\mu|=0.8\), respectively. In this setting, the weak target is easily masked by the residual response and IQI-induced image of the strong target. The proposed elliptic GOMP provides a much steeper increase in \(P_D\) of both targets and reaches reliable recovery at lower SNR than the reference receivers. This gain is particularly pronounced under strong IQI in Fig.~3(b), where the compensation-first and OFDM-ZF-based methods remain unreliable, while the proposed detector approaches a high $P_D$. This confirms that removing the strong target through its coupled direct-plus-mirror elliptic response exposes nearby weak targets more effectively.
% Copy the preamble lines below into your LaTeX document preamble if needed.
% \usepackage{pgfplots}
% \pgfplotsset{compat=1.18}
% \usepgfplotslibrary{groupplots}

% ==================== fig1_snr_sweep.tex ====================
% Requires in the main preamble: \usepackage[caption=false,font=footnotesize]{subfig}, \usepackage{pgfplots}, and \pgfplotsset{compat=1.18}
\begin{figure}[t]
\centering
\subfloat[$K=10$, 16QAM, $|\lambda/\mu|=0.5$]{%
\begin{tikzpicture}
\begin{axis}[
width=0.8\columnwidth,
height=0.58\columnwidth,
grid=both,
xlabel={SNR (dB)},
ylabel={$P_D$},
ymin=0, ymax=1.02,
xmin=-20, xmax=20,
legend style={font=\tiny, at={(0.52,0.02)}, anchor=south west},
legend cell align={left}
]
\addplot+[thick, mark=o] table[x index=0, y index=1] {fig1_snr_sweep_rho_0p5_stdOfdmZfExact.txt};
\addplot+[thick, mark=o] table[x index=0, y index=1] {fig1_snr_sweep_rho_0p5_pairwiseIqZfExact.txt};
\addplot+[thick, mark=o] table[x index=0, y index=1] {fig1_snr_sweep_rho_0p5_genieLmmseExact.txt};
\addplot+[thick, mark=o] table[x index=0, y index=1] {fig1_snr_sweep_rho_0p5_ellipticGompExact.txt};
\legend{OFDM-ZF+OMP,Pairwise IQI-ZF+OMP, LMMSE+OMP,Elliptic GOMP}
\end{axis}
\end{tikzpicture}
}\\[-0.5ex]
\subfloat[$K=10$, 16QAM, $|\lambda/\mu|=0.8$]{%
\begin{tikzpicture}
\begin{axis}[
width=0.8\columnwidth,
height=0.58\columnwidth,
grid=both,
xlabel={SNR (dB)},
ylabel={$P_D$},
ymin=0, ymax=1.02,
xmin=-20, xmax=20,
legend style={draw=none},
legend cell align={left}
]
\addplot+[thick, mark=o] table[x index=0, y index=1] {fig1_snr_sweep_rho_0p8_stdOfdmZfExact.txt};
\addplot+[thick, mark=o] table[x index=0, y index=1] {fig1_snr_sweep_rho_0p8_pairwiseIqZfExact.txt};
\addplot+[thick, mark=o] table[x index=0, y index=1] {fig1_snr_sweep_rho_0p8_genieLmmseExact.txt};
\addplot+[thick, mark=o] table[x index=0, y index=1] {fig1_snr_sweep_rho_0p8_ellipticGompExact.txt};
\end{axis}
\end{tikzpicture}
}%
\caption{$P_D$ versus SNR for different receiver IQI strengths.}
\label{fig:fig1-snr-sweep}
\end{figure}
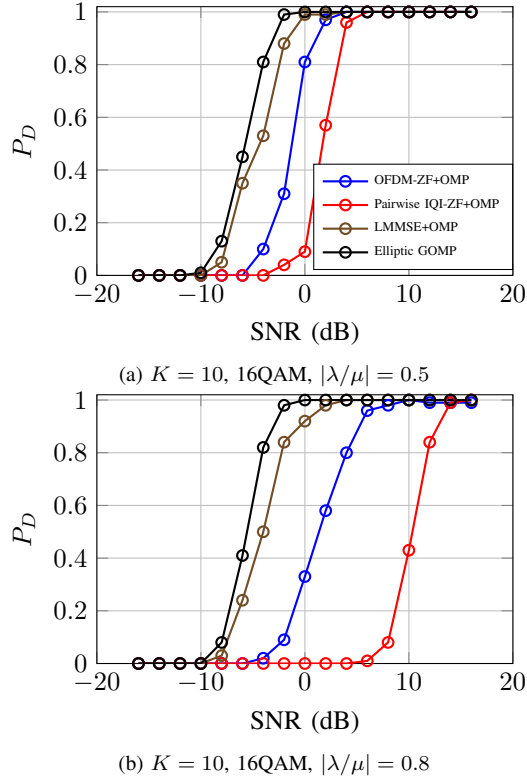

% ==================== fig2_sparsity_sweep.tex ====================
% Requires in the main preamble: \usepackage[caption=false,font=footnotesize]{subfig}, \usepackage{pgfplots}, and \pgfplotsset{compat=1.18}
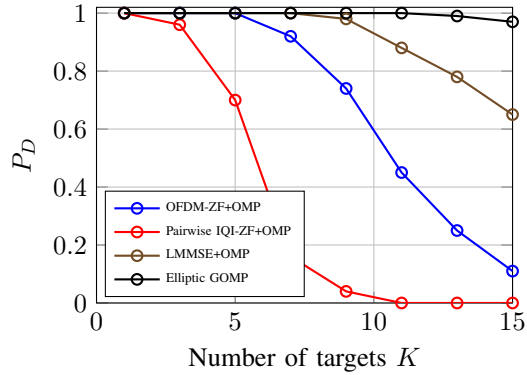
\begin{figure}[t]
\centering
\begin{tikzpicture}
\begin{axis}[
width=0.8\columnwidth,
height=0.62\columnwidth,
grid=both,
xlabel={Number of targets $K$},
ylabel={$P_D$},
ymin=0, ymax=1.02,
xmin=0, xmax=15,
legend style={font=\tiny, at={(0.02,0.02)}, anchor=south west},
legend cell align={left}
]
\addplot+[thick, mark=o] table[x index=0, y index=1] {fig2_sparsity_sweep_K_stdOfdmZfExact.txt};
\addplot+[thick, mark=o] table[x index=0, y index=1] {fig2_sparsity_sweep_K_pairwiseIqZfExact.txt};
\addplot+[thick, mark=o] table[x index=0, y index=1] {fig2_sparsity_sweep_K_genieLmmseExact.txt};
\addplot+[thick, mark=o] table[x index=0, y index=1] {fig2_sparsity_sweep_K_ellipticGompExact.txt};
\legend{OFDM-ZF+OMP,Pairwise IQI-ZF+OMP, LMMSE+OMP,Elliptic GOMP}
\end{axis}
\end{tikzpicture}
\caption{$P_D$ versus number of targets}
\label{fig:fig2-sparsity-sweep}
\end{figure}

% ==================== fig3_near_strong_sweep.tex ====================
% Requires in the main preamble: \usepackage[caption=false,font=footnotesize]{subfig}, \usepackage{pgfplots}, and \pgfplotsset{compat=1.18}
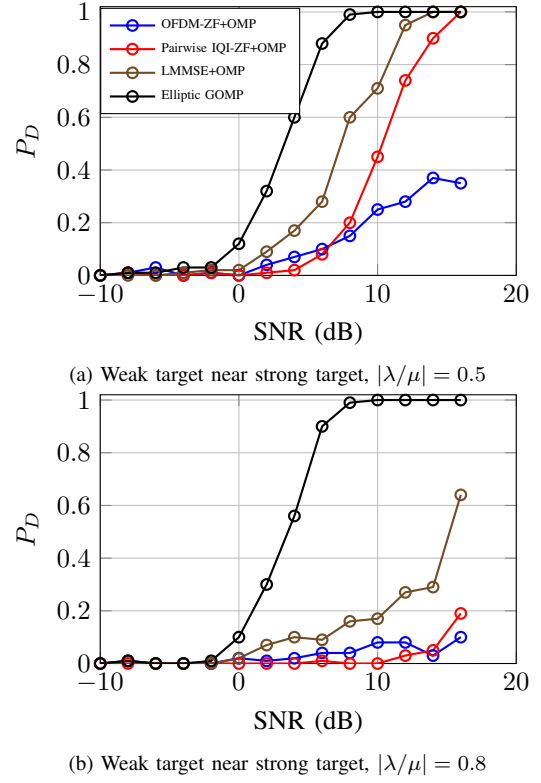
\begin{figure}[t]
\centering
\subfloat[Weak target near strong target, $|\lambda/\mu|=0.5$]{%
\begin{tikzpicture}
\begin{axis}[
width=0.8\columnwidth,
height=0.58\columnwidth,
grid=both,
xlabel={SNR (dB)},
ylabel={$P_D$},
ymin=0, ymax=1.02,
xmin=-10, xmax=20,
xmin=-10,
legend style={font=\tiny, at={(0,0.6)}, anchor=south west},
legend cell align={left}
]
\addplot+[thick, mark=o] table[x index=0, y index=1] {fig3_near_strong_sweep_near_rho_0p5_stdOfdmZfExact.txt};
\addplot+[thick, mark=o] table[x index=0, y index=1] {fig3_near_strong_sweep_near_rho_0p5_pairwiseIqZfExact.txt};
\addplot+[thick, mark=o] table[x index=0, y index=1] {fig3_near_strong_sweep_near_rho_0p5_genieLmmseExact.txt};
\addplot+[thick, mark=o] table[x index=0, y index=1] {fig3_near_strong_sweep_near_rho_0p5_ellipticGompExact.txt};
\legend{OFDM-ZF+OMP,Pairwise IQI-ZF+OMP, LMMSE+OMP,Elliptic GOMP}
\end{axis}
\end{tikzpicture}
}\\[-0.5ex]
\subfloat[Weak target near strong target, $|\lambda/\mu|=0.8$]{%
\begin{tikzpicture}
\begin{axis}[
width=0.8\columnwidth,
height=0.58\columnwidth,
grid=both,
xlabel={SNR (dB)},
ylabel={$P_D$},
ymin=0, ymax=1.02,
xmin=-10, xmax=20,
legend style={draw=none},
legend cell align={left}
]
\addplot+[thick, mark=o] table[x index=0, y index=1] {fig3_near_strong_sweep_near_rho_0p8_stdOfdmZfExact.txt};
\addplot+[thick, mark=o] table[x index=0, y index=1] {fig3_near_strong_sweep_near_rho_0p8_pairwiseIqZfExact.txt};
\addplot+[thick, mark=o] table[x index=0, y index=1] {fig3_near_strong_sweep_near_rho_0p8_genieLmmseExact.txt};
\addplot+[thick, mark=o] table[x index=0, y index=1] {fig3_near_strong_sweep_near_rho_0p8_ellipticGompExact.txt};
\end{axis}
\end{tikzpicture}
}%
\caption{Weak target recovery for different receiver IQI.}
\label{fig:fig3-near-strong-sweep}
\end{figure}

\vspace{-0.3cm}
\section{Conclusion}
This letter introduced an elliptic range-Doppler processing framework for OFDM-ISAC under receiver IQI. We showed that receiver IQI transforms each physical target into a coupled direct-plus-mirror response, which is represented by an elliptic atom group rather than a single Fourier atom. The proposed elliptic GOMP detector operates directly on the IQI-impaired OFDM grid and computes the required correlations through two weighted two-dimensional FFTs, preserving the leading FFT order of conventional sparse processing. Numerical results demonstrated improved exact support recovery, robustness to increasing target density, and reliable weak-target detection near strong scatterers. These gains confirm that the IQI-induced mirror branch can be utilized as a useful sensing structure rather than treated only as an impairment.
\bibliographystyle{IEEEtran}
\bibliography{bibliography}

@ARTICLE{isac_1,
  author={Liu, Fan and Cui, Yuanhao and Masouros, Christos and Xu, Jie and Han, Tony Xiao and Eldar, Yonina C. and Buzzi, Stefano},
  journal={IEEE J. Sel. Areas Commun.}, 
  title={{Integrated Sensing and Communications: Toward Dual-Functional Wireless Networks for 6G and Beyond}}, 
  year={2022},
  volume={40},
  number={6},
  pages={1728-1767},
  doi={10.1109/JSAC.2022.3156632}}

@ARTICLE{isac_2,
  author={Zhang, J. Andrew and Liu, Fan and Masouros, Christos and Heath, Robert W. and Feng, Zhiyong and Zheng, Le and Petropulu, Athina},
  journal={IEEE J. Sel. Topics Signal Process.}, 
  title={{An Overview of Signal Processing Techniques for Joint Communication and Radar Sensing}}, 
  year={2021},
  volume={15},
  number={6},
  pages={1295-1315},
  doi={10.1109/JSTSP.2021.3113120}}

@ARTICLE{isac_3,
  author={Sturm, Christian and Wiesbeck, Werner},
  journal={Proc. IEEE}, 
  title={{Waveform Design and Signal Processing Aspects for Fusion of Wireless Communications and Radar Sensing}},
  year={2011},
  volume={99},
  number={7},
  pages={1236-1259},
  doi={10.1109/JPROC.2011.2131110}}

@inproceedings{sota_1,
  title={{OFDM Radar Algorithms in Mobile Communication Networks}},
  author={K. Braun},
  year={2014},
  url={https://api.semanticscholar.org/CorpusID:69276211}
}

@INPROCEEDINGS{sota_2,
  author={Bourdoux, André and Bauduin, Marc and Desset, Claude},
  booktitle={Proc. IEEE Radar Conference (RadarConf)}, 
  title={{IQ Imbalance Robust and Low PAPR OFDM Radar Waveform}}, 
  year={2019},
  volume={},
  number={},
  pages={1-6},
  doi={10.1109/RADAR.2019.8835496}}

@ARTICLE{sota_3,
  author={Lang, Oliver and Hofbauer, Christian and Tockner, Moritz and Feger, Reinhard and Wagner, Thomas and Huemer, Mario},
  journal={IEEE Trans. Veh. Technol.}, 
  title={{OFDM-Based Waveforms for Joint Sensing and Communications Robust to Frequency Selective IQ Imbalance}}, 
  year={2025},
  volume={74},
  number={1},
  pages={1078-1091},
  doi={10.1109/TVT.2024.3463802}}

@article{sota_4,
  author  = {Meingassner, Andreas and Lang, Oliver and Tockner, Moritz and Plaimer, Bernhard and Wagner, Matthias and Lindorfer, G{\"u}nther and Hofstadler, Michael and Huemer, Mario},
  title   = {{Tx and Rx IQ imbalance compensation for JCAS in 5G NR}},
  journal = {Journal on Advances in Signal Processing},
  year    = {2026},
  volume  = {2026},
  number  = {52},
  pages   = {52},
  doi     = {10.1186/s13634-026-01329-9},
}

@ARTICLE{sota_5,
  author={Zheng, Le and Wang, Xiaodong},
  journal={IEEE Trans. Signal Process.}, 
  title={{Super-Resolution Delay-Doppler Estimation for OFDM Passive Radar}}, 
  year={2017},
  volume={65},
  number={9},
  pages={2197-2210},
  doi={10.1109/TSP.2017.2659650}}

@ARTICLE{sota_6,
  author={Li, Peishi and Li, Ming and Liu, Rang and Liu, Qian and Lee Swindlehurst, A.},
  journal={IEEE Trans. Wireless Commun.}, 
  title={{MIMO-OFDM ISAC Waveform Design for Range-Doppler Sidelobe Suppression}}, 
  year={2025},
  volume={24},
  number={2},
  pages={1001-1015},
  doi={10.1109/TWC.2024.3503605}}

@ARTICLE{sota_7,
  author={Liu, Fan and Zhang, Ying and Xiong, Yifeng and Li, Shuangyang and Yuan, Weijie and Gao, Feifei and Jin, Shi and Caire, Giuseppe},
  journal={IEEE Trans. Inf. Theory}, 
  title={{CP-OFDM Achieves the Lowest Average Ranging Sidelobe Under QAM/PSK Constellations}}, 
  year={2025},
  volume={71},
  number={9},
  pages={6950-6967},
  doi={10.1109/TIT.2025.3591267}}

@ARTICLE{rx_iqi,
  author={Tarighat, A. and Bagheri, R. and Sayed, A.H.},
  journal={IEEE Transactions on Signal Processing}, 
  title={Compensation schemes and performance analysis of IQ imbalances in OFDM receivers}, 
  year={2005},
  volume={53},
  number={8},
  pages={3257-3268},
  doi={10.1109/TSP.2005.851156}}
\end{document}